# Reflection-free finite volume Maxwell's solver for adaptive grids

Nina Elkina[1,*], Hartmut Ruhl

*Ludwig-Maximilians Universität München, 80539, Germany*

**Abstract**

We present a non-staggered method for the Maxwell equations in adaptively refined grids. The code is based on finite volume central scheme that preserves in a discrete form both divergence-free property of magnetic field and the Gauss law. High spatial accuracy is achieved with help of non-oscillatory extrema preserving piece-wise or piece-wise-quadratic reconstructions. The semi-discrete equations are solved by implicit-explicit Runge-Kutta method. The new adaptive grid Maxwell's solver is examined based on several 1d examples, including the an propagation of a Gaussian pulse through vacuum and partially ionised gas. Two-dimensional extension is tested with a Gaussian pulse incident on dielectric disc. Additionally, we focus on testing computational accuracy and efficiency.

## 1. Introduction

Adaptive mesh refinement (AMR) [1] allows to increase grid resolution locally when and where it is required physically relevant shortest length scale of problem under consideration. This makes AMR a highly desirable technique for kinetic simulation of multi-scale relativistic effects in plasma [2]. Our interest to AMR is motivated by recent studies of electron-positron cascades driven by ultra-strong laser fields [3], [4]. Next generation of laser facility will allow to study such processes in the laboratory [5]. Quantum cascades are very localized phenomenon which may cause stiffness of coupled Maxwell's and kinetic plasma equations. The aim of this paper is to develop an adaptive mesh solver for the Maxwell equations in a way which allows to treat a general class of localized and stiff problems.

[*]Corresponding author
 *Email address:* `Nina.Elkina@physik.uni-muenchen.de` (Nina Elkina)

Often the case that numerical schemes show spurious reflections when applied to adaptive (non-uniform) meshes. These artefacts can be observed as wave packet propagating through change in grid resolution splits into transmitted and reflected parts. Then these reflections can be trapped in regions of fine grid resolution or even lead to numerical instability. Reflections were also observed in Maxwell's solvers [6] and [7] which are based on a popular in plasma simulations Yee's scheme [8]. In central scheme numerical artefacts are often attributed to a non-monotonic character of grid dispersion which supports modes with different sign of the group velocity [9]. Decoupled in uniform mesh these modes can couple across the grid interface. Though dispersion of Yee's scheme is monotonic the grid staggering artificially couples distinct branches at grid interfaces [10]. Effect of spurious reflection can be mitigated to some extent by smoothing the transition between grids [9] or by high-order solution interpolation across the interface [7]. Reflections caused by high frequency modes can be reduced by smoothing of electric and magnetic fields as suggested in [11] for staggered scheme. In this paper we propose to exploit a natural dissipation properties of explicit finite volume method (FV) with complex dispersion law. Such schemes damps out most reflection-affected high-frequency modes automatically. Also conservative cell-centred FV methods [12] can be easily extended to any type of grids including adaptively refinement ones. FV discretization of the Maxwell equation have been presented in [13] where authors proposed a divergence-corrected Godunov's method based on solution of Riemann problem. In this paper we adopt an alternative Riemann-free semi-discrete central scheme proposed in [14]. This class of methods also allows for un-split integration of stiff source terms using an appropriate time stepping routine [15]. A core part of central FV schemes is a polynomial reconstruction of the solution from cell averages computed at the previous time step. Reconstructed solution is then used to calculate numerical fluxes. Stability of FV schemes heavily relies on monotonic (non-oscillatory) property of the reconstruction enforced with the help of limiters [16]. Roughly speaking the limiting simply reverts the reconstruction to monotonic first order near discontinuities [17]. Limiters are not always good in recognition of smooth extrema where limiting is not needed. Such unjustified limiting can be prevented with help of additional extrema indicators acting on top of limiters for detection of smooth minima or maxima [18]. An essential ingredient of any Maxwell's solver is maintaining of divergence constraints imposed by the Gauss law $\nabla E = 4\pi\rho$ and divergence-free property of magnetic field, i.e. $\nabla \cdot B = 0$. Violation of divergence conditions causes severe numerical problems like non-conservation of charge in plasma [19]. Accumulated divergence errors can be efficiently cleaned out by introducing a generalized Lagrange multiplier [20]. Application of this technique results in corrected Maxwell equations with additional terms responsible for transport of divergence errors. However for the central finite volume method is more preferable to keep an original form of the Maxwell equations on discrete level. For this reason we adopt an alternative approach based on an appropriate redistribution of numerical fluxes [21] formulated in terms of vertex-based auxiliary potentials [22].

The reminder of this paper will be organized as follows. Section 2 and Section 3 describe the central finite volume method and time stepping routine. High order solution reconstructions with extension to adaptive meshes is discussed in Section 4. In Section 5 we compare FV-AMR solver with one based on Yee's method and discuss grid dispersion properties. The Maxwell system is extended to nonlinear problem related to a tunnel ionization of plasma in Section 6 Multi-dimension version of the code is tested in Section 7 by simulating the 2d Gaussian pulse incident onto refractive disc. Our conclusions are summarized in Section 8.

## 2. Basic equations and numerical scheme

In the following we will numerically integrate evolutionary part of the Maxwell system

$$\frac{1}{c}\frac{\partial \boldsymbol{E}}{\partial t} = \nabla \times \boldsymbol{B} - \frac{4\pi}{c}\boldsymbol{J}, \tag{1}$$

$$-\frac{1}{c}\frac{\partial \boldsymbol{B}}{\partial t} = \nabla \times \boldsymbol{E}, \tag{2}$$

$$\tag{3}$$

where $\boldsymbol{E}$, $\boldsymbol{B}$ and $\boldsymbol{J}$ are the electric, magnetic fields and current density respectively. Remaining pair of the Maxwell equations

$$\nabla \cdot \boldsymbol{E} = 4\pi\rho, \quad \nabla \cdot \boldsymbol{B} = 0, \tag{4}$$

serves a role of constraints imposed on electric and magnetic field which provide continuity equation in plasma

$$\frac{\partial \rho}{\partial t} + \nabla \cdot \boldsymbol{J} = 0, \tag{5}$$

where $\rho$ is the charge density.

Let us cast the Maxwell equation in form of conservation law system

$$\frac{\partial \boldsymbol{q}}{\partial t} + \frac{\partial f^x(\boldsymbol{q})}{\partial x} + \frac{\partial f^y(\boldsymbol{q})}{\partial y} + \frac{\partial f^z(\boldsymbol{q})}{\partial z} = \boldsymbol{S}(\boldsymbol{q}), \tag{6}$$

where $\boldsymbol{q}$ is the vector of conserved variables, $\boldsymbol{\mathcal{F}}(\boldsymbol{q}) = \{f^x(\boldsymbol{q}), f^y(\boldsymbol{q}), f^z(\boldsymbol{q})\}$ is the flux functions and $\boldsymbol{S}(\boldsymbol{q})$ is the source terms. A detailed transcription on variables reads

$$\begin{aligned}
\boldsymbol{q} &= \{E_x, & E_y, & E_z, & B_x, & B_y, & B_z\} \\
f^x(\boldsymbol{q}) &= \{0, & B_z, & -B_y, & 0, & -E_z, & E_y\} \\
f^y(\boldsymbol{q}) &= \{-B_z, & 0, & B_x, & E_z, & 0, & -E_x\} \\
f^z(\boldsymbol{q}) &= \{B_y, & -B_x, & 0, & -E_y, & E_x, & 0\} \\
\boldsymbol{S}(\boldsymbol{q}) &= 4\pi \{J_x/c, & J_y/c, & J_z/c, & 0, & 0, & 0\}
\end{aligned} \tag{7}$$

A usual rout to derive finite volume method [12] is to integrate a conservation law over a cell $\Omega_{i,j} = \Delta x \Delta y$ as follows

$$\int_\Omega \frac{\partial \boldsymbol{q}}{\partial t} + \int_\Omega \nabla \cdot \boldsymbol{\mathcal{F}} = \int_\Omega S d\Omega, \tag{8}$$

Applying the Gauss theorem in order to replace volume integral by the contour one over cell edges results in equation for the cell average

$$\frac{\partial \overline{q}_{i,j}}{\partial t} + \frac{1}{\Omega_{i,j}} \oint_{\partial\Omega_{i,j}} \boldsymbol{\mathcal{F}} \cdot \boldsymbol{n} dC = S_{i,j}, \tag{9}$$

where $\partial\Omega_{i,j}$ denotes the surface of control volume, and $\boldsymbol{n}$ is the outward vector. The cell average is defined as

$$\overline{q}_{i,j} = \frac{1}{\Delta x_{i,j} \Delta y_{i,j}} \int_{x_{i-1/2,j}}^{x_{i+1/2,j}} \int_{y_{i,j-1/2}}^{y_{i,j+1/2}} q(x,y) dx dy. \tag{10}$$

Calculating the contour integral in (9) with mid-point rule we arrive at semi-discrete finite volume scheme

$$\frac{d\overline{q}_{i,j}}{dt} = -\frac{\mathcal{F}_{i+1/2,j} - \mathcal{F}_{i-1/2,j}}{\Delta x} - \frac{\mathcal{G}_{i,j+1/2} - \mathcal{G}_{i,j-1/2}}{\Delta y}, \tag{11}$$

where numerical fluxes are approximated with the two-point Lax-Wendroff function

$$\mathcal{F}_{i+1/2,j}(q^+, q^-) = \frac{a^x}{2}\left[f^x(q^-_{i+1/2,j}) + f^x(q^+_{i+1/2,j})\right] - \frac{a^x}{2}\left(q^+_{i+1/2,j} - q^-_{i+1/2,j}\right), \tag{12}$$

$$\mathcal{G}_{i,j+1/2}(q^+, q^-) = \frac{a^y}{2}\left[f^y(q^-_{i,j+1/2}) + f^y(q^+_{i,j+1/2})\right] - \frac{a^y}{2}\left(q^+_{i,j+1/2} - q^-_{i,j+1/2}\right). \tag{13}$$

As input parameters this function takes two values $q^+$ and $q^-$ of reconstructed solution computed at both sides of cell edge (for more details about reconstruction see Section 4)

$$q^+_{i+1/2,j} = q_{i+1,j}(x_{i+1/2}), \quad q^-_{i+1/2,j} = q_{i+1,j}(x_{i-1/2}), \tag{14}$$

where $\boldsymbol{a}$ is local speed of solution estimated using eigenvalues of the Jacobian $\boldsymbol{a} = \max(\lambda_k\{\partial_q \boldsymbol{\mathcal{F}}(\boldsymbol{q})\}, 0) = \pm c$.

In order to satisfy divergence conditions (4) in discrete equations (11) we adopt auxiliary potentials method [22]. Consider a minimal setting for the transverse mode

$$\frac{1}{c}\frac{d}{dt}\begin{pmatrix} B_x \\ B_y \\ E_z \end{pmatrix}_{i,j} = \frac{1}{\Delta x}\begin{pmatrix} 0 \\ E_z \\ -B_y \end{pmatrix} + \frac{1}{\Delta y}\begin{pmatrix} -E_z \\ 0 \\ B_x \end{pmatrix}. \tag{15}$$

The corresponding conserved variables and flux functions are given by

$$q = \{B_x, B_y, E_z\}, \quad \mathcal{F} = \{0, E_z, -B_y\}, \quad \mathcal{G} = \{-E_z, 0, B_x\}. \qquad (16)$$

We define vertex-based potentials as follows

$$A^1_{i+1/2,j+1/2} = \frac{1}{2}\left(\mathcal{G}^1_{i,j+1/2} + \mathcal{G}^1_{i+1,j+1/2}\right) \qquad\qquad = \mu_x \mathcal{G}_{ij}, \qquad (17)$$

$$A^2_{i+1/2,j+1/2} = \frac{1}{2}\left(\mathcal{F}^2_{i+1/2,j} + \mathcal{F}^2_{i+1/2,j+1}\right) \qquad\qquad = \mu_y \mathcal{F}_{ij}, \qquad (18)$$

$$A^3_{i+1/2,j+1/2} = \frac{1}{4}\left(\mathcal{F}^3_{i+1/2,j} + \mathcal{F}^3_{i+1/2,j+1} + \mathcal{G}^3_{i,j+1/2} + \mathcal{G}^3_{i+1,j+1/2}\right) = \frac{1}{2}\mu_y \mathcal{F}_{ij} + \frac{1}{2}\mu_x \mathcal{G}_{ij}, \qquad (19)$$

New fluxes then are given by averaging of two vertex potentials over cell corner

$$\boldsymbol{\mathcal{F}}^*_{i+1/2,j} = \mu_x \boldsymbol{A}_{i,j+1/2}, \quad \boldsymbol{\mathcal{G}}^*_{i,j+1/2} = \mu_y \boldsymbol{A}_{i+1/2,j}, \qquad (20)$$

Numerical schemes now reads

$$\frac{1}{c}\frac{d}{dt}\begin{pmatrix} B^x \\ B^y \\ E^z \end{pmatrix}_{i,j} = \frac{\delta^x \boldsymbol{\mathcal{F}}^*_{i,j}}{\Delta x} + \frac{\delta^y \boldsymbol{\mathcal{G}}^*_{i,j}}{\Delta y} = \frac{1}{\Delta x}\begin{pmatrix} 0 \\ \delta^x \mu^y A^3 \\ \delta^x \mu^y A^2 \end{pmatrix} + \frac{1}{\Delta y}\begin{pmatrix} \delta^x \mu_x A^3 \\ 0 \\ -\delta^y \mu^x y A^1 \end{pmatrix}, \qquad (21)$$

where $\delta$ is the difference operator, i.e.

$$\delta^x \boldsymbol{\mathcal{F}}^*_{i,j} = (\boldsymbol{\mathcal{F}}^*_{i+1/2,j} - \boldsymbol{\mathcal{F}}^*_{i-1/2,j})/2, \quad \delta^y \boldsymbol{\mathcal{G}}^*_{i,j} = (\boldsymbol{\mathcal{G}}^*_{i,j+1/2} - \boldsymbol{\mathcal{G}}^*_{i,j-1/2})/2.$$

Preservation of $\nabla \cdot \boldsymbol{B} = 0$ can be easily proven for modified scheme

$$\frac{d}{dt}(\nabla \cdot \boldsymbol{B})_{i,j} = \frac{d}{dt}\left[\frac{\partial B^x}{\partial x} + \frac{\partial B^y}{\partial y}\right]_{i,j} \qquad (22)$$

$$= \frac{d}{dt}\left(\frac{1}{\Delta x}\mu^y \delta^x B^x_{i,j} + \frac{1}{\Delta y}\mu^x \delta^y B^y_{i,j}\right) = \frac{1}{\Delta x \Delta y}\left[-\mu^y \delta^x \delta^y \mu^x + \mu^x \delta^y \delta^x \mu^y\right] A^3_{i,j} = 0. \qquad (23)$$

Auxiliary potentials provide a transverse correction over vertexes for electric and magnetic fields which naturally incorporates divergence conditions in the scheme.

### 3. Time-stepping with IMEX Runge-Kutta

Semi-discrete formulation of the scheme (11) leaves gives a freedom in selection of suitable time stepping routine to solve an ordinary differential equation(ODE)

$$\frac{dq}{dt} = \mathcal{F}(q) + \mathcal{S}(q). \tag{24}$$

The Runge-Kutta type methods are described by the following set of stage equations

$$K^i = q^n + \Delta t \sum_{j=1}^{i-1} \tilde{a}_{ij}\mathcal{F}_q(t + c_j\Delta h, Q_i) + \Delta t \sum_{j+1}^{\nu} a_{i,j}\mathcal{S}(t_n + c_{i,j}\Delta t,\ Q_j), \tag{25}$$

$$q^{n+1} = q^n + \Delta t \sum_{i=1}^{\nu} \tilde{w}_i \mathcal{F}_q(t_n + \tilde{c}_i \Delta t,\ Q_i) + \Delta t \sum_{i=1}^{\nu} w_i \mathcal{S}(t_n + c_i \Delta h,\ Q_i), \tag{26}$$

are often implemented to solve semi-discrete equations. In this paper we consider a mixed explicit-implicit (IMEX) Runge-Kutta SSP(3,3,2) scheme defined by the double Butcher's table

$$
\begin{array}{c|ccc}
0 & 0 & 0 & 0 \\
1/2 & 1/2 & 0 & 0 \\
1 & 1/2 & 1/2 & 0 \\
\hline
a & 1/3 & 1/3 & 1/3
\end{array}
\qquad
\begin{array}{c|ccc}
1/4 & 1/4 & 0 & 0 \\
1/4 & 0 & 1/4 & 0 \\
1 & 1/3 & 1/3 & 1/3 \\
\hline
\tilde{a} & 1/3 & 1/3 & 1/3
\end{array}
\tag{27}
$$

A detailed transcription of scheme (26-27) reads

$$q^{(0)} = q^n \qquad\qquad\qquad\qquad\qquad\qquad\qquad\qquad + \frac{1}{4}\Delta\Big[\mathcal{S}(q^{(0)})\Big], \tag{28}$$

$$q^{(1)} = q^n + \frac{1}{2}\Delta t\Big(\mathcal{F}(q^{(0)})\Big) \qquad\qquad\qquad\qquad + \frac{1}{4}\Delta t\Big[\mathcal{S}(q^{(1)})\Big], \tag{29}$$

$$q^{(2)} = q^n + \frac{1}{2}\Delta t\Big(\mathcal{F}(q^{(0)}) + \mathcal{F}(q^{(1)})\Big) \qquad + \frac{1}{3}\Delta t\Big[\mathcal{S}(q^{(0)}) + \mathcal{S}(q^{(1)}) + \mathcal{S}(q^{(2)})\Big], \tag{30}$$

$$q^{n+1} = q^n + \frac{1}{3}\Delta t\bigg(\mathcal{F}(q^{(0)}) + \mathcal{F}(q^{(1)}) + \mathcal{F}(q^{(2)})\bigg) + \frac{1}{3}\Delta t\bigg[\mathcal{S}(q^{(0)}) + \mathcal{S}(q^{(1)}) + \mathcal{S}(q^{(2)})\bigg]. \tag{31}$$

In absence of the source term this scheme reduces to a fully explicit strong-stability preserving Runge-Kutta method [23]. We apply fixed point iteration to solve this non-linear system.

## 4. Reconstruction and adaptive grid

As it was noted in the previous Section t flux functions need point-wise values of solution taken from both sides of cell edges. Solution can be obtained using a simple piece-wise reconstruction

$$\boldsymbol{q}_{i,j}(x,y) = \overline{\boldsymbol{q}}_{i,j} + D^x_{i,j} \cdot (x - x_i) + D^y_{i,j} \cdot (y - y_j), \tag{32}$$

where $D^x$ and $D^y$ are the discrete derivatives approximated by centred differences. In smooth regions this reconstruction provides a solution of second order spatial accuracy. In non-smooth regions additional care has to be taken about monotonicity of the reconstruction. This can be done by calculation of discrete derivatives using for example MinMod limiter

$$D^x(q, \theta) = \mathrm{MinMod}\left(\theta \frac{q_i - q_{i-1}}{\Delta x_i/2 + \Delta x_{i-1}/2}, \frac{q_{i+1} - q_{i-1}}{\Delta x_{i+1}/2 + \Delta x_i + \Delta x_{i+1}/2}, \theta \frac{q_{i+1} - q_i}{\Delta x_{i+1} + \Delta x_i/2}\right), \tag{33}$$

where the MinMod limiter is defined as

$$\mathrm{MinMod}(z_1, z_2, ..., z_j) = \begin{cases} \min_j & \text{if } z_j > 0, \quad \forall j \\ \max_j & \text{if } z_j < 0, \quad \forall j \\ 0 & \text{otherwise} \end{cases}, \tag{34}$$

is employed with $1 \leq \theta \leq 2$. This limiter is a subject of extrema clipping effects which leads to lower-order solution representation around smooth minima or maxima. The accuracy can be restored with additional indicator for recognition of smooth extrema where limiting is not needed. In our code we have implemented an extrema-preserving limiter proposed in [18] which is shortly described in Appendix Appendix B.

An accurate approximation of solution beyond second order can be achieved with higher-reconstructions. One of successful example is to combine both piecewise and quadratic polynomials as suggested in [24]

$$q_i(x) = (1 - \theta_i) \cdot \mathcal{L}_i(x) + \theta_i \cdot \mathcal{Q}_i(x), \tag{35}$$

where $\theta_i$ is limiter which blends different reconstructions depending on local smoothness of solution. Quadratic polynomial $\mathcal{Q}_i(x)$ is defined by

$$\mathcal{Q}_i(x) = A + B \cdot (x - x_i) + C \cdot (x - x_i)^2. \tag{36}$$

Where coefficients $A$, $B$ and $C$ can be uniquely defined from conservation of solution over cell

$$\frac{1}{\Delta x_i} \int_{x_i - \Delta x_i/2}^{x_i + \Delta x/2} \mathcal{Q}(x')dx' = \bar{q}_i, \quad j = i-1, i, i-1. \tag{37}$$

For uniform grid the coefficients are defined by

$$A = -\frac{1}{24}\left[q_1 - 26q_0 + 2q_{-1}\right], \quad B = \frac{1}{60\Delta x}\left[37q_1 - 14q_0 - 23q_{-1}\right], \quad C = \frac{2}{5\Delta x^2}\left[q_1 - 2q_0 + q_{-1}\right]. \tag{38}$$

The limiter $\theta_i$ in equation (35) can be calculated as follows

$$\theta_i = \begin{cases} \min\left[\frac{M_{i+1/2} - L_{i+1/2}}{M_i - L_{i+1/2}}, \frac{m_{i-1/2} - L_{i-1/2}}{m_i - L_{i-1/2}}\right], & \text{if } \bar{q}_{i-1} < \bar{q}_i < \bar{q}_{i+1}, \\ \min\left[\frac{M_{i-1/2} - L_{i-1/2}}{M_i - L_{i-1/2}}, \frac{m_{i+1/2} - L_{i+1/2}}{m_i - L_{i+1/2}}\right], & \text{if } \bar{q}_{i-1} > \bar{q}_i > \bar{q}_{i+1}, \\ 1, & \text{otherwise}, \end{cases} \tag{39}$$

where $L_{i\pm 1/2} = L_i(x_{i\pm 1/2})$. Variables $M$ and $m$ are defined as

$$M_i = \max(\mathcal{Q}_i(x_{i+1/2}), \mathcal{Q}_i(x_{i-1/2})), \tag{40}$$

$$m_i = \min(\mathcal{Q}_i(x_{i+1/2}), \mathcal{Q}_i(x_{i-1/2})), \tag{41}$$

and

$$M_{i\pm 1/2} = \max\left\{\frac{1}{2}L_i(x_{i\pm 1/2}) + L_{i\pm 1}(x_{i\pm 1/2}), \mathcal{Q}_{i\pm 1}(x_{i\pm 1/2})\right\}, \tag{42}$$

$$m_{i\pm 1/2} = \min\left\{\frac{1}{2}L_i(x_{i\pm 1/2}) + L_{i\pm 1}(x_{i\pm 1/2}), \mathcal{Q}_{i\pm 1}(x_{i\pm 1/2})\right\}. \tag{43}$$

Extension to adaptive grid is especially easy in finite volume method since only reconstruction step has to be modified. Communication between cells happens at reconstruction stage when cell averages from neighbour cell are requested for computing discrete derivatives. This communication is organized with help of interface functions which provide information on required resolution level, either coarsened, refined or left unaffected. If refinement ratio between adjacent cell is fixed by factor of at most these transfer functions are especially simple. For coarsening and refining we apply the same reconstruction polynomial as described above. For example from coarse level to finer one is given by

$$q^{l+1} = \frac{4}{V} \int_V [q^l + D_{ij}^{x,l}(x - x_i) + D^{y,l}(y - y_j)] dV,$$

Setting the $x_i = 0$, $y_j = 0$ (cell centre), one can immediately obtain

$$q^{l+1} = \frac{4}{\Delta x \Delta y} \left[ q_{ij}^l \frac{xy}{4} + D_{ij}^{x,l} \frac{x^2 y}{8} + D_{ij}^{y,l} \frac{y^2 x}{8} \right] \quad (44)$$

In the opposite case of transfer from finer $q^l$ to coarse grid $q^{l-1}$ the integration of reconstruction polynomial reduces to summation over finer cells

$$q_{i,j}^l = \frac{1}{4}(q_{2i,2j}^{l+1} + q_{2i,2j+1}^{l+1} + q_{2i+1,2j}^{l+1} + q_{2i+1,2j+1}^{l+1}). \quad (45)$$

This can be efficiently optimized if cell of the same refinement level are organized in blocks and communication happens between such blocks. The exchange of information is realized as a local boundary conditions defined for each block. Local boundaries consist of 3 layers of ghost cells which dynamically filled up with data obtained from neighbour blocks with an appropriate refining/coarsening procedure.

## 5. 1D Gaussian pulse on adaptive grid

Let us first illustrate how grid reflections look like with example Yee's scheme. For this reason we consider analytically of a Gaussian pulse propagating in vacuum

$$\boldsymbol{A} = a_0\{0, \cos(kx - t\omega_0), \sin(\phi)\} \times \exp\left(-\frac{(kx - t\omega_0)^2}{2\sigma^2}\right), \quad (46)$$

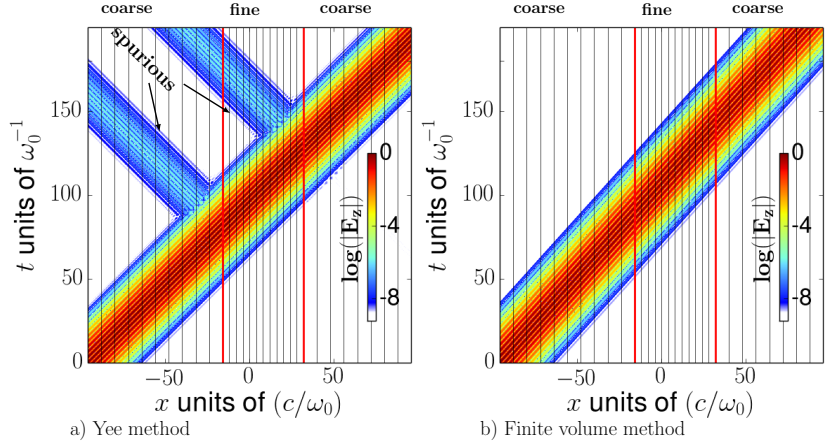

Figure 1: Space-time diagram of laser pulse propagation $\log(|\boldsymbol{E}(x,t)|)$ through refinement grid patch. Yee method (a) exhibits spuriously reflecting waves at both coarse-fine and fine-coarse interfaces. The results obtained with finite element method (b) are reflection free.

where $\boldsymbol{A}$ is the vector potential, $\sigma$ is the half-width-maximum, $a_0$ and $\omega_0$ are the laser amplitude and frequency. Numerical solution obtained with the Yee scheme is presented in Figure 1 (left). Here pulse initialized at $t = 0$ using equation (46) propagates through simple mesh with one level of refinement. Reflection appears while pulse goes across the each grid refinement level. The same initial conditions integrated with FV method do not show any sign of reflection (see right panel). Let us now consider dispersion properties of both schemes in 1d case (see Appendix Appendix A for details)

$$\omega_{\text{Yee}} = \pm \frac{2}{\Delta x} \sin\left(\frac{k\Delta x}{2}\right), \tag{47}$$

$$\omega_{\text{fv}} = \frac{i}{2\Delta x}\left[4\sin^2\left(\frac{k\Delta x}{2}\right) - \sin^2(k\Delta x) \pm i\sin(k\Delta x)[3 - \cos(k\Delta x)]\right]. \tag{48}$$

Dispersion law of the Yee scheme is real for all frequencies $\omega = \omega_r + i \cdot 0$. Thus the Yee scheme propagates all modes without being damped. The FV scheme demonstrate non-monotonic form of dispersion curve peaked at $k\Delta x = 0.6\pi$.

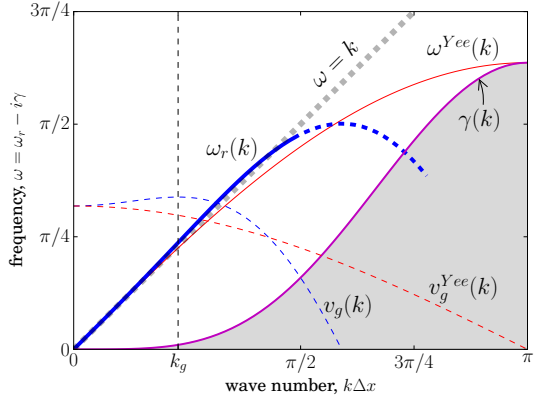

Figure 2: (Color online) Dispersion relation for finite volume method and the Yee scheme. Real frequency $\omega_r$ (solid blue line) and imaginary $\gamma$ (filled grey region) damping coefficient. versus wave number $k$. The Yee's scheme propagates solution without damping (i.e. $\gamma_{Yee} = 0$).

Group velocities for both methods $v_g = \partial \omega / \partial k$ are given by

$$v_g^{FV} = \frac{3}{2}\cos(k\Delta x) - \frac{1}{2}\cos(2k\Delta x), \quad v_g^{Yee} = \cos\left(\frac{k\Delta x}{2}\right). \qquad (49)$$

Group velocity changes sign for waves with $k\Delta x > 0.6\pi$. These modes are subject or reflection across grid interface due to artificial coupling a dispersion branch which supports backward propagating waves [9] and [10]. These modes are effectively damped out provided by imaginary part of frequency $\omega = \omega_r + i\gamma$, $\gamma > 1$. The decrement of damping is shown in Figure with grey shaded area. It can be seen that dissipation gets noticeable already at $k > k_g$, where $k_g = \arccos(3/4) \cdot \Delta x^{-1}$. However grid resolved modes in range However we most interested to resolve waves $(0, k_g)$ remains almost unaffected by damping. As one can see in Figure, wave modes below $k_g$ are propagated faster then speed of light in FV scheme while Yee's one decelerate all modes. This property of FV method is especially beneficial for relativistic plasma simulation since it prevents spurious resonance between particles in waves.

We proceed with study of numerical accuracy measured using $L_2$ norm as

$$L_2 = \frac{1}{N}\left(\sum |q_i - q_i^{ex}|^2\right)^{1/2}, \qquad (50)$$

|      | FV2 |  |  |  | FV3 |  |  |  |
|------|-------------|-------|------------|-------|--------------|------|------------|------|
| Base | *Single level* | | *Two levels* | | Single level | | Two Levels | |
|      | Error | Order | Error | Order | Error | Rate | Error | Rate |
| 128  | 0.094206 |      | 0.098428 |      | 0.46  |      | 0.374 |      |
| 256  | 0.038847 | 2.42 | 0.038734 | 2.54 | 0.13  | 3.53 | 0.106 | 3.53 |
| 512  | 0.014073 | 2.76 | 0.014084 | 2.75 | 0.035 | 3.7  | 0.027 | 3.59 |

Table 1: Error and convergence rate in $L_2$.

where $\Delta x = L_x/N_x$ and $N_x$ is the number of grid cells. Exact solution $q^{ex}$ is given by (46). For adaptive grid this definition should be slightly modified as follows [25]

$$L_2^{amr} = \left( \frac{\sum^N \Omega_i (q_i - q^{ex})^2}{\sum^N \Omega_i} \right)^{1/2}, \qquad (51)$$

where $\Omega_i$ is the volume of cell $i$ (in 1d case $\Omega_i = \Delta x_i$). We collect results for piece-wise and piece-wise-quadratic reconstructions in Table 1. Each type of reconstruction demonstrates higher order then expected from formal one, i.e. super-convergence. Refinement is done in form presented in Figure 1, a base grid which contains one or two levels of refinement. One can see that second level of refinement does not always improve accuracy. We attribute this effect to over-resolution when discrete precision errors proportional to a number of grid cell and time steps accumulates faster then decrease of scheme error with grid refinement.

### 6. Tunnel ionization by laser pulse

In order to test the ability of our solver to handle non-linear source term we carry out simulation of dynamic ionization by laser field. I The model for

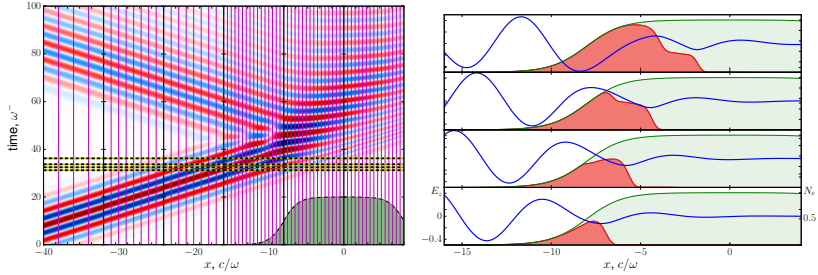

Figure 3: (left) Space-time distribution of laser field $E_z$ incident on neutral gas layer (shaded with green). (right) Snapshots of electron density (red shaded) at time moments (denoted by dashed lines in left panel) with corresponding $E_z(x)$ profiles (continuous lines).

initially neutral gas is adopted from [26]

$$\frac{\partial N'_e}{\partial t'} = \frac{w(|\bm{E}'|)}{w_0}(N'_g - N'_e), \tag{52}$$

$$\frac{\partial \bm{J}}{\partial t} = N'_e \bm{E}', \tag{53}$$

$$w(|\bm{E}'|) = 4\omega_a \left(\frac{I_\alpha}{I_h}\right)^{5/2} \frac{1}{|\bm{E}'|} \exp\left[-\frac{2}{3}\left(\frac{I_\alpha}{I_h}\right)^{3/2} \frac{1}{|\bm{E}|}\right]. \tag{54}$$

This model describes generation of new electrons with density $N_e$ from initially neutral gas $N_g(x)$. Here we apply the following normalization

$$E = E' E_a, \quad t = t' \omega_0, \quad N_e = N'_e N_{cr}, \quad \text{where} \quad N_{cr} = \omega_0^2 m / 4\pi e^2, \tag{55}$$

where $N_{cr}$ is the critical plasma density, $E_a = m^2 e^5/\hbar^4 = 5.1 \times 10^9$ V/cm is the field strength at the Bohr radius. Tunnel ionization is governed by probability $w(|\bm{E}|)$ as given in [27]

$$w(|\bm{E}|) = 4\omega_a \left(\frac{I_\alpha}{I_h}\right)^{5/2} \frac{E_\alpha}{|\bm{E}|} \exp\left[-\frac{2}{3}\left(\frac{I_\alpha}{I_h}\right)^{3/2} \frac{E_\alpha}{|\bm{E}|}\right], \tag{56}$$

where $I_a$ and $I_h$ are the ionization potentials of atom and hydrogen respectively (here $I_a/I_h = 1$), $\omega_a = me^4/\hbar^3 = 4.16 \times 10^{16}$ s$^{-1}$ is the atom frequency. Initial profile of neutral gas is defined by

$$N_g = N_0 \cdot (\tanh[0.5(x + L_g/2)] - \tanh[0.5(x - L_g/2)])/2, \tag{57}$$

where $L_g = 16c/\omega_0$ is the characteristic size of gas layer, $N_0 = 1$. In order to resolve gas layer we make use of a grid with 2 refined levels with finest resolution concentrated at neutral gas profile depicted with shadowed region in Figure 3. Time step used in simulation is $\Delta t = 0.25 \cdot \Delta x^0/2^{l-1}$, where $\Delta x^0 = 0.25 \cdot c/\omega_0^{-1}$ is the initial refinement level. Incident laser pulse comes from the left in simulation box $L_x = [-64c/\omega_0,\ 64c/\omega_0]$ propagates though subsequently refined grids without reflections and hits gas layer. Ionization process is illustrated in left part of Figure where a few snapshots of electron density are shown. A newly created plasma is dense enough to reflect back laser pulse at later stages of simulation.

## 7. Two-dimensional simulation: Interaction with refractive disc

Two-dimensional tests are carried out for a problem of laser pulse incident on dielectric refractive disc [7]. Electromagnetic effects in medium are taken into account in the following form of the Maxwell equations

$$\nabla(\varepsilon \boldsymbol{E}) = 4\pi\rho, \quad \nabla \cdot \boldsymbol{B} = 0 \tag{58}$$

$$-\frac{1}{c}\frac{\partial \boldsymbol{B}}{\partial t} = \nabla \times \boldsymbol{E} \tag{59}$$

$$\frac{1}{c}\frac{\partial(\varepsilon \boldsymbol{E})}{\partial t} = \nabla \times \left(\frac{\boldsymbol{B}}{\mu}\right) - \frac{4\pi}{c}\boldsymbol{J} \tag{60}$$

$$\tag{61}$$

where solved in a box with 3 refinement levels. It was assumed vacuum magnetic permeability $\mu = 1$, the dielectric susceptibility $\varepsilon$ depends on refractive index $n = \sqrt{\mu\varepsilon}$ given by

$$n = \begin{cases} 2, & x^2 + y^2 < R^2 \\ 1, & \text{otherwise.} \end{cases} \tag{62}$$

The refractive disc of radius $R = 2$ is placed at $x = 0$, $y = 0$. The laser pulse is initialized with the Gaussian profile

$$\boldsymbol{E} = a_0 \cdot \boldsymbol{p} \cdot \cos(\boldsymbol{k} \cdot \boldsymbol{r}) \cdot \text{Env} \tag{63}$$

$$\boldsymbol{B} = a_0 \cdot (\boldsymbol{k} \times \boldsymbol{p}) \cdot \cos(\boldsymbol{k} \cdot \boldsymbol{r}) \cdot \text{Env}, \tag{64}$$

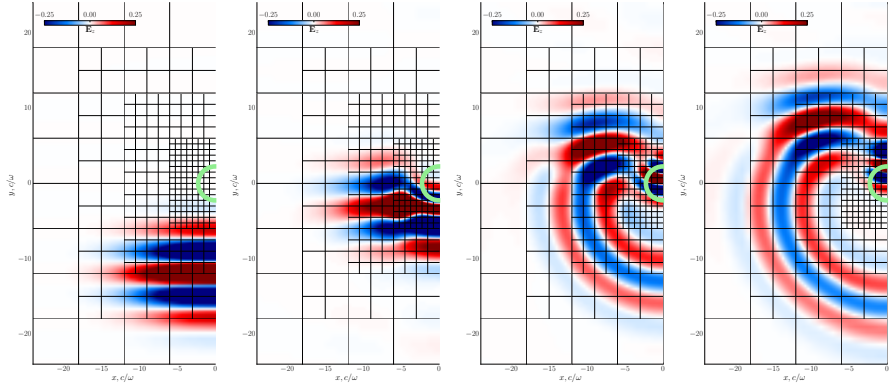

Figure 4: Time evolution of Gaussian pulse incident on a refractive disc.

where $\boldsymbol{k}$, $\boldsymbol{p}$ are the wave and polarization vectors respectively and the envelop is defined by

$$\mathrm{Env}(\boldsymbol{r}_\parallel,\ \boldsymbol{r}_\perp) = \exp\left[-\frac{r_\parallel^2}{\sigma_\parallel^2} - \frac{r_\perp^2}{\sigma_\perp^2}\right]$$

where $r_\parallel = \boldsymbol{k} \cdot \boldsymbol{r}$ and $r_\perp = |\boldsymbol{r} \times \boldsymbol{k}|$ are the parallel and perpendicular components of radius vector $\boldsymbol{r}$ with respect to wave vector $\boldsymbol{k}$. The half-width-maximum $\sigma_\perp = \sigma_\parallel$. The linearly polarized pulse $\boldsymbol{p}0, 0, 1$ pulse propagates along $y$-directions $\boldsymbol{k} = 0, 1, 0$. A sequence of snapshots of $E_z(x)$ is plotted in Figure 4. These picture shows smooth and stable propagation of a pulse though refractive disc. The grid in this simulation contains 4 different refinement levels with peak resolution $512 \times 512$ which covers a disc. The coarsest grid level is $64 \times 64$ can be found near boundaries of the box. In order to measure the accuracy of this case we use a well-resolved 'reference case' solution on the grid $512 \times 512$ taken at time $T = 18\omega_0^{-1}$. The time step for uniform grid is $\Delta t = 0.25h$, where $h = \Delta x = \Delta y$. In adaptive grid simulation we define global time step as $\Delta t = 0.25(\Delta x_0/2^{l-1})$, where $l$ is a number of refinement levels used in simulation. One can see that AMR time step depends on second finest level and therefore the time step on the most refined grid is larger compared with globally uniform grid simulations. The reference solution was re-projected to a corresponding coarser level by means equation (44). All tested configuration and corresponding $L_2$ norms are presented in Figure 5, left panel. Observed accuracy of simulations is at least of second order as it can be seen from comparison with a pair of helper dashed lines showing 1st and 2nd order slopes. Accuracy improves towards 3rd order if initial refinement level increases. This super-convergence effect we have observed with 1d case as well. We deduce a relative complexity for presented simulation runs in Figure 5, left. In this plot we normalize the complexity (number of ) of each simulation is normalized by a reference case $N_{512 \times 512} = N_{\text{cell}} \times N_{\text{step}} = 512^2 \times 768$. For quantitative estimates let's consider a medium uniform resolution $128 \times 128$ case which is 64 times faster then reference case. Two-level refined grid (marked with diamonds) gives acceleration about $100 times$, while finer one-level grid (square markers) gives only factor of 50 but with much better accuracy. Thus a gain in computational performance is considerable however we expect much better results after implementation of adaptive time stepping and dynamic grid adaptivity.

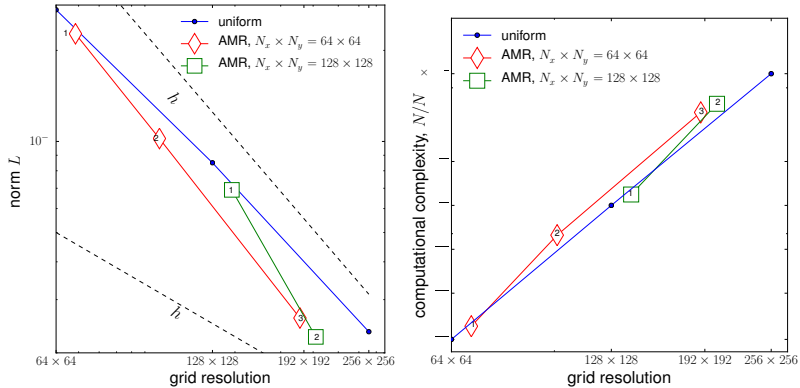

Figure 5: (left) $L_2$ as a function of grid points calculated using uniform grid reference case with resolution $512 \times 512$. A number of refinement levels for AMR cases are shown inside markers. (right) Relative complexity of computations normalized by the reference case.

## 8. Conclusions

A new Maxwell solver has proven to be very effective for adaptive mesh computation. The spurious reflections are eliminated by means of natural dissipation property of central finite volume method. The damping depends on wave mode increasing at most reflection-affected lager wave numbers which strongly deviate from physical dispersion. High accuracy of the code is achieved using piece-wise or blended piece-wise-quadratic reconstruction in conjunction with extrema preserving limiter. The implicit-explicit time stepping makes our code applicable for simulation of stiff problems. Numerical study of convergence with $L_2$ norm shown that code demonstrate good accuracy and even super-convergence effect detected for higher grid resolutions. This new solver has then been tested with study of Gaussian beam propagation in various media resolved with adapted grids. By direct comparison with the Yee solver we demonstrate that our code does not show spurious reflections while former one does. The ability of code to deal with non-linear plasma effects has been demonstrated by modelling of dynamic ionization by laser pulse propagating in neutral gas. For this problem we apply adaptive mesh to resolve a localized layer of dense gas. After laser hits gas layer ionization creates plasma. A newly created plasma eventually reflects laser pulse back. We perform also two dimensional simulation of a laser pulse incident on 2d refractive disc. We observed a noticeable growth of computational performance compared with uniformly resolved case. We expect further improvement of efficiency after implementation of adaptive time stepping. The results resented in this paper are very encouraging and give strong motivation for further improvement of our code by making it dynamically adaptive in both space and time. Our future plans are to incorporate our AMR-Maxwell solver into particle-in-cell and multi-fluid plasma simulation codes.

**Acknowledgement**

The research is partially supported by the DFG TR18/B13 and DFG RU 633/1-1, the Cluster-of-Excellence "Munich Center for Advanced Photonics" (MAP) and Arnold Sommerfeld Zentrum (ASZ).

**Appendix A. Dispersion properties of finite volume method**

Let us we consider an impact of numerical dispersion effects for simplest one dimensional case

$$\frac{1}{c}\frac{\partial E_z}{\partial t} = \frac{\partial B_y}{\partial x}, \quad \frac{1}{c}\frac{\partial B_y}{\partial t} = \frac{\partial E_z}{\partial x}. \tag{A.1}$$

Semi-discretized form of this equations reads

$$\dot{\bar{B}} = \frac{c}{\Delta x}[E_{i+1/2} - E_{i-1/2}], \quad \dot{\bar{E}} = \frac{c}{\Delta x}[B_{i+1/2} - B_{i-1/2}]. \tag{A.2}$$

We seek for solution of this equation in form $q = \hat{q}\exp(ikx + i\omega t)$, where $q = \{E, B\}$. Thus

$$q^+_{i+1/2} = \bar{q}_i + \frac{\bar{q}_{i+1} - \bar{q}_{i-1}}{4}, \quad \hat{q}^+_{i+1/2} = \hat{q}e^{ikx_i}a, \tag{A.3}$$

$$q^-_{i+1/2} = \bar{q}_{i+1} - \frac{\bar{q}_{i+2} - \bar{q}_i}{4}, \quad \hat{q}^-_{i+1/2} = \hat{q}e^{ikx_i}e^{ik\Delta x}b, \tag{A.4}$$

where $a = 1 + i\sin(k\Delta x)/2$ and $b = 1 - i\sin(k\Delta x)/2$. Thus Fourier transformed fluxes (A.2) take form

$$\hat{F}_{i+1/2} = \frac{\hat{E}_z}{2}A(a + b \cdot e^{+ik\Delta x}) - \frac{\hat{B}_y}{2}(a - be^{+ik\Delta x}), \tag{A.5}$$

$$\hat{F}_{i-1/2} = \frac{\hat{E}_z}{2}(b + a \cdot e^{-ik\Delta x}) - \frac{\hat{B}_y}{2}(ae^{-ik\Delta x} - b). \tag{A.6}$$

where $A$ and $B$

$$A = a(1 - e^{-ik\Delta x}) - b(1 - e^{ik\Delta x}) = i\sin(k\Delta x)[3 - \cos(k\Delta x)], \tag{A.7}$$

$$B = a(1 - e^{-ik\Delta x}) + b(1 - e^{ik\Delta x}) = 4\sin^2\left(\frac{k\Delta x}{2}\right) - \sin^2(k\Delta x). \tag{A.8}$$

$$\begin{pmatrix} -2hi\omega - K & -M \\ -M & -2hi\omega - K \end{pmatrix} \begin{bmatrix} \hat{E} \\ \hat{B} \end{bmatrix} = 0 \tag{A.9}$$

The complex valued solution reads

$$\omega = \frac{i}{2\Delta x}(B \pm A). \tag{A.10}$$

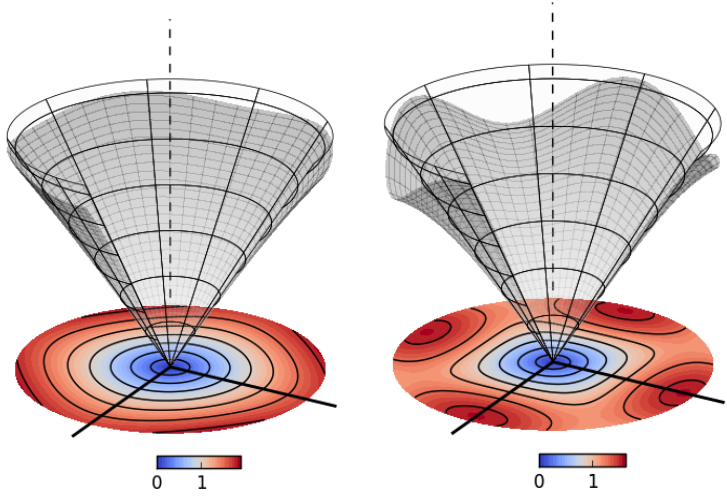

Figure A.6: Two dimensional dispersion relation for Yee's (left panel) and FV (right panel) schemes.

The Figure A.6 shows real part of frequency for FV and Yee's schemes plotted as a function go wave number $\boldsymbol{k}$.

$$\hat{G}_{i,j+1/2} = \frac{\hat{E}_z}{2}[a_y + e^{+k_y \Delta y} \cdot b_y] - \frac{\hat{B}_x}{2}[a_y - e^{+k_y \Delta y} \cdot b_y], \tag{A.11}$$

$$\hat{\psi}_{i+1/2,j+1/2} = \frac{\hat{E}_z}{4}\left([1 + e^{k_x \Delta x}][a_y + b_y e^{+k_y \Delta y}]\right) - \frac{\hat{B}_x}{4}\left([1 + e^{k_x \Delta x}][a_y - b_y e^{+k_y \Delta y}]\right), \tag{A.12}$$

$$G^*_{i,j+1/2} = \frac{\hat{E}_z}{4}\left[1 + \cos(k_x \Delta x)\{a_y + b_y e^{+k_y \Delta y}\}\right] - \frac{\hat{B}_x}{4}\left[1 + \cos(k_x \Delta x)\{a_y - b_y e^{+k_y \Delta y}\}\right] \tag{A.13}$$

where

$$a_x, b_x = 1 \pm i \sin(k_x \Delta x), \quad a_y, b_y = 1 \pm i \sin(k_y \Delta y).$$

These potential-based fluxes go to Fourier transformed scheme (11)

$$i\omega B_x = -\frac{\hat{E}_z}{4\Delta y}A + \frac{\hat{B}_x}{4\Delta y}B, \tag{A.14}$$

$$i\omega B_y = +\frac{\hat{E}_z}{4\Delta x}C - \frac{\hat{B}_y}{4\Delta x}D, \tag{A.15}$$

$$i\omega \hat{E}_z = +\frac{1}{4\Delta y}\left[\hat{B}_x A - \hat{E}_z B\right] - \frac{1}{4\Delta x}\left[\hat{B}_y C - \hat{E}_z D\right] \tag{A.16}$$

where $A, B, C, D$ are given by

$$A = (1 + \cos(k_x \Delta x))[(1 - e^{-ik_y \Delta y})a_y - (1 - e^{ik_y \Delta y})b_y], \tag{A.17}$$

$$B = (1 + \cos(k_x \Delta x))[(1 - e^{-ik_y \Delta y})a_y + (1 - e^{ik_y \Delta y})b_y], \tag{A.18}$$

$$C = (1 + \cos(k_y \Delta y))[(1 - e^{-ik_x \Delta x})a_x - (1 - e^{ik_x \Delta x})b_x], \tag{A.19}$$

$$D = (1 + \cos(k_y \Delta y))[(1 - e^{-ik_x \Delta x})a_x + (1 - e^{ik_x \Delta x})b_x]. \tag{A.20}$$

The dispersion of FV scheme takes a form of cubic equation

$$a\omega^3 + b\omega^2 + c\omega + d = 0, \tag{A.21}$$

where

$$a = -64i, \tag{A.22}$$

$$b = 32(B - D), \tag{A.23}$$

$$c = -4i(3BD + A^2 + C^2 - D^2 - B^2), \tag{A.24}$$

$$d = BC^2 + DB^2 - BD^2 - DA^2. \tag{A.25}$$

In order to compare FV dispersion law with one corresponding to the Yee scheme in 2d

$$\omega_{Yee} = \sqrt{2\sin(k_x\Delta x/2)^2 + 2\sin(k_y\Delta y/2)^2}, \tag{A.26}$$

we plot both equations in Figure A.6. One can see that the Yee scheme provides more isotropic picture of wave propagation. However FV scheme can be easily improved in this regard by adding the cross derivatives terms in polynomial reconstruction.

**Appendix B. Multidimensional extrema preserving limiter**

For the sake of completeness we provide a description of 2d extrema preserving limiter [18] needed to compute $D^x_{i,j}$ and $D^y_{i,j}$ in piece-wise linear reconstruction 32.

The monotonic property states that local reconstruction must be bounded by cell averages from neighbours cells, i.e.

$$\min(\boldsymbol{Q}_{i,j}) < q_{i,j}(x,y) < \max(\boldsymbol{Q}_{i,j}). \tag{B.1}$$

we define $\boldsymbol{Q}_{i,j}$ by

$$\overline{\boldsymbol{Q}} = \begin{bmatrix} \overline{q}_{i-1,j+1} & \overline{q}_{i,j+1} & \overline{q}_{i+1,j+1} \\ \overline{q}_{i-1,j} & \overline{q}_{i,j} & \overline{q}_{i+1,j} \\ \overline{q}_{i-1,j-1} & \overline{q}_{i,j-1} & \overline{q}_{i+1,j-1} \end{bmatrix}, \tag{B.2}$$

and corresponding derivatives are given by

$$\boldsymbol{D}^x_\pm = \begin{bmatrix} (\overline{q}_{i-1,j+1} - \overline{q}_{i,j}) & (\overline{q}_{i,j+1} - \overline{q}_{i,j}) & (\overline{q}_{i+1,j+1} - \overline{q}_{i,j}) \\ (\overline{q}_{i-1,j} - \overline{q}_{i,j}) & \pm\epsilon & (\overline{q}_{i+1,j} - \overline{q}_{i+1,j}) \\ (\overline{q}_{i-1,j-1} - \overline{q}_{i,j}) & (\overline{q}_{i,j-1} - \overline{q}_{i,j-1}) & (\overline{q}_{i+1,j-1} - \overline{q}_{i+1,j}) \end{bmatrix}, \tag{B.3}$$

where constant $\epsilon = 10^{-20}$ enforces continuous dependence on data. Monotonic condition for reconstruction $q(x,y)$ reads

$$\boldsymbol{V}_{\min} < \partial q(x,y) < \boldsymbol{V}_{\max}, \tag{B.4}$$

where $\boldsymbol{V}_{\min} = \min(\boldsymbol{D}^-)$ and $\boldsymbol{V}_{\max} = \max(\boldsymbol{D}^+)$ are minimum and maximum of local derivatives. Slope that preserves monotonicity can be defined by adjusting this inequality at cell corners as follows. Point-wise solution at upper right corner $q_3 = q_{i+1/2,j+1/2}$ is given by

$$q_3 = q_{i,j} + D^x/2 + D^y/2, \tag{B.5}$$

$$\partial q_3 = (q_3 - \overline{q}_{i,j}) = D^x/2 + D^y/2 \tag{B.6}$$

where $D^x$ and $D^y$ are uncorrected central slopes

$$D^x = \frac{q_{i+1,j} - q_{i-1,j}}{2}, \quad D^y = \frac{q_{i,j+1} - q_{i,j-1}}{2}. \tag{B.7}$$

In order to put $\partial q_3$ into allowed interval (B.4) one can define a weighting factor

$$w_3 = \text{Weight}(\partial q_3, V_{\min}, V_{\max}) = \begin{cases} V_{\max}/\partial q_3, & \partial q_3 > V_{\max}, \\ V_{\min}/\partial q_3, & \partial q_3 < V_{\min}, \\ 1, & \text{otherwise} \end{cases} \quad (B.8)$$

After all weights attributed to each corners calculated corrected slopes are computed as

$$\mathcal{D}^x_{i,j} = \min(w_1, w_2, w_3, w_4) \cdot D^x_{i,j}, \quad \mathcal{D}^y_{i,j} = \min(w_1, w_2, w_3, w_4) \cdot D^y_{i,j}. \quad (B.9)$$

These slopes is then used in reconstruction. This multi-dimensional limiter still can suffer of extrema clipping. Unnecessary limiting can be avoided if we consider data along diagonal direction $q_{i+p,j+p}$, where $p = -1, 2$. Looking in extended interval $[\bar{q}_{i-1,j-1}, \bar{q}_{i+2,j+2}]$ we can judge whether or not smaller interval $[\bar{q}_{i,j}, \bar{q}_{i+1,j+1}]$ falls outside. If so then we near smooth extrema then allowed slopes can be found in larger interval

$$[V_{\min}, V_{\max}] \to [V_{\min} - \epsilon_2 \cdot d_1, V_{\max} + \epsilon_2 \cdot d_1], \quad (B.10)$$

where $\epsilon_2 = 0.275$ is the empirical constant and coefficient $d_1$ discriminates extrema ($d_1 \neq 0$) and discontinuities ($d_1 = 0$). Parameter $d_1$ is calculated as following

$$d_1 = \max(t_{\max} - w_{\max}, w_{\min} - t_{\min}, 0), \quad (B.11)$$

where

$$t_{\max} = \max(\bar{q}_{i,j}, \bar{q}_{i+1,j+1}), \quad (B.12)$$
$$t_{\min} = \min(\bar{q}_{i,j}, \bar{q}_{i+1,j+1}), \quad (B.13)$$
$$w_{\max} = \max(\bar{q}_{i-1,j-1}, \bar{q}_{i+2,j+2}), \quad (B.14)$$
$$w_{\min} = \min(\bar{q}_{i-1,j-1}, \bar{q}_{i+2,j+2}), \quad (B.15)$$
$$(B.16)$$

[1] M. J. Berger, J. Oliger, Adaptive Mesh Refinement for Hyperbolic Partial Differential Equations, Journal of Computational Physics 53 (1984) 484.

[2] C. Birdsall, A. Langdon, Plasma Physics via Computer Simulation, Series in Plasma Physics and Fluid Dynamics, Taylor & Francis, 2004. URL `https://books.google.de/books?id=S2lqgDTm6a4C`

[3] N. V. Elkina, A. M. Fedotov, I. Y. Kostyukov, M. V. Legkov, N. B. Narozhny, E. N. Nerush, H. Ruhl, QED cascades induced by circularly polarized laser fields, Physical Review Special Topics Accelerators and Beams 14 (5) (2011) 054401. `arXiv:arXiv:1010.4528, doi:10.1103/PhysRevSTAB.14.054401`.

[4] E. N. Nerush, I. Y. Kostyukov, A. M. Fedotov, N. B. Narozhny, N. V. Elkina, H. Ruhl, Laser Field Absorption in Self-Generated Electron-Positron Pair Plasma, Physical Review Letters 106 (3) (2011) 035001. `doi:10.1103/PhysRevLett.106.035001`.

[5] G. A. Mourou, T. Tajima, S. V. Bulanov, Optics in the relativistic regime, Reviews of Modern Physics 78 (2006) 309–371. `doi:10.1103/RevModPhys.78.309`.

[6] J.-L. Vay, An Extended FDTD Scheme for the Wave Equation: Application to Multiscale Electromagnetic Simulation, Journal of Computational Physics 167 (2001) 72–98.

[7] A. R. Zakharian, M. Brio, C. Dineen, J. V. Moloney, Second-order accurate FDTD space and time grid refinement method in three space dimensions, IEEE Photonics Technology Letters 18 (2006) 1237–1239.

[8] K. Yee, Numerical solution of initial boundary value problems involving Maxwell's equations in isotropic media, IEEE Transactions on Antennas and Propagation 14 (1966) 302–307.

[9] R. Vichnetsky, Propagation through numerical mesh refinement for hyperbolic equations, Mathematics and Computers in Simulation (1981) 344–353.

[10] J. Frank, S. Reich, Energy-conserving semi-discretizations and spurious numerical reflections, Tech. Rep. MAS-E0608, Centrum voor Wiskunde en Informatica, P.O. Box 94079, 1090 GB Amsterdam (Oct. 2004).

[11] K. Fujimoto, A new electromagnetic particle-in-cell model with adaptive mesh refinement for high-performance parallel computation, Journal of Computational Physics 230 (2011) 8508–8526. `doi:10.1016/j.jcp.2011.08.002`.

[12] R. Leveque, Finite-Volume Methods for Hyperbolic Problems, Cambridge University Press, Cambridge, UK, 2002.

[13] C.-D. Munz, P. Ommes, R. Schneider, A three-dimensional finite-volume solver for the maxwell equations with divergence cleaning on unstructured meshes, Computer Physics Communications 130 (12) (2000) 83 – 117. `doi:http://dx.doi.org/10.1016/S0010-4655(00)00045-X`.
URL `http://www.sciencedirect.com/science/article/pii/S001046550000045X`

[14] K. Kurganov, S. Noelle, G. Petrova, Semi-discrete central-upwind schemes for hyperbolic conservation laws and Hamilton-Jacobi equations, SIAM J. Sci. Comput. 23 (2001) 707–740.

[15] U. Asher, R. Ruuth, S. Spiteri, Implicit-explicit Runge-Kutta method for time dependent Partial Differential Equations, Appl. Numer. Math. 25 (4) (1997) 151–167.

[16] P. Sweby, High resolution schemes using flux limiters for hyperbolic conservation laws, SIAM J. Numer. Anal. 21 (1984) 995–1011.

[17] S. K. Godunov, A difference method for numerical calculation of discontinuous solutions of the equations of hydrodynamics, Matematicheskii Sbornik 89 (3) (1959) 271–306.

[18] A. Suresh, Positivity-Preserving Scheme in Multidimensions, SIAM J. Sci. Comput. 22 (4) (2000) 1184–1198.

[19] C. Munz, R. Schneider, E. Sonnendrucker, U. Voss, Maxwell's equations when the charge conservation is not satisfied, Academie des Sciences Paris Comptes Rendus Serie Sciences Mathematiques 328 (1999) 431–436. `doi: 10.1016/S0764-4442(99)80185-2`.

[20] C.-D. Munz, P. Omnes, R. Schneider, E. Sonnendrücker, U. Voß, Divergence Correction Techniques for Maxwell Solvers Based on a Hyperbolic Model, Journal of Computational Physics 161 (2000) 484–511. `doi: 10.1006/jcph.2000.6507`.

[21] M. Torrilhon, M. Fey, Constraint preserving upwind methods for multidimensional advection equations, SIAM J. Numer. Anal. 42 (2004) 1694–1728.

[22] S. Mishra, E. Tadmor, Constraint preserving schemes using potential-based fluxes. III. Genuinely multidimensional schemes for MHD equations, ESAIM: M2AN 46 (2012) 661–680.

[23] S. Gottlieb, On high order strong stability preserving Runge-Kutta and Multi Step Time Discretizations, Journal of Scientific Computing 25 (2005) 105–128.

[24] X. D. L. Liu, S. Osher, Non-oscillatory high-order accurate self-similar maximum principle satisfying shock capturing schemes I, SIAM 33 (2) (1996) 760–779. `doi:10.1103/PhysRevLett.106.035001`.

[25] M. Sun, K. Takayama, Error localisation in solution-adaptive grid methods, Comp. Phys. 190 (2003) 346–350.

[26] E. S. Efimenko, A. V. Kim, M. Quiroga-Teixeiro, Ionization-Induced Small-Scaled Plasma Structures in Tightly Focused Ultrashort Laser Pulses, Physical Review Letters 102 (2009) 015002. `doi:10.1103/PhysRevLett.106.035001`.

[27] B. M. Karnakov, V. D. Mur, S. V. Popruzhenko, V. S. Popov, Current progress in developing the nonlinear ionization theory of atoms and ions,